\documentclass[12pt]{article}
\usepackage{amssymb,latexsym}
\usepackage[cmex10]{amsmath}
\usepackage[all]{xy}
\input{xy}
\xyoption{all}
\title{A Note on the Games-Chan Algorithm}
\author{Graham H. Norton
\footnote{School of Mathematics and Physics, University of Queensland, Brisbane, Queensland 4072, Australia. (Email: ghn@maths.uq.edu.au)}
}

\newtheorem{theorem}{\bf Theorem}[section]
\newtheorem{corollary}[theorem]{\bf Corollary}
\newtheorem{proposition}[theorem]{\bf Proposition} 
\newtheorem{notation}[theorem]{\sc Notation} 
\newtheorem{definition}[theorem]{\bf Definition} 
\newtheorem{lemma}[theorem]{\bf Lemma} 
\newtheorem{example}[theorem]{\it Example}  
\newtheorem{examples}[theorem]{\it Examples}  
\newtheorem{algorithm}[theorem]{\bf Algorithm} 
\newtheorem{conjecture}[theorem]{\bf  Conjecture}
\newtheorem{remark}[theorem]{\it Remark}
\newtheorem{remarks}[theorem]{\it  Remarks}

\newenvironment{proof}{{\noindent\it Proof.\ }}{$\square$\par\vspace{4mm}} 
\def \bt{ \begin{theorem} }
\def \et{ \end  {theorem} }
\def \bl{ \begin{lemma} }
\def \el{ \end  {lemma} }
\def \bp{ \begin{proposition} }
\def \ep{ \end  {proposition} }
\def \bn{ \begin{notation} }
\def \en{ \end  {notation} }
\def \bq{ \begin {question} }
\def \eq{ \end {question} }
\def \bc{ \begin{corollary} }
\def \ec{ \end  {corollary} }
\def \bcj{ \begin{conjecture} }
\def \ecj{ \end  {conjecture} }
\def \bd{ \begin{definition} }
\def \ed{ \end  {definition} }
\def \bdp{ \begin{definitionprop} }
\def \edp{ \end  {definitionprop} }
\def \bdt{ \begin{definitiontheorem} }
\def \edt{ \end  {definitiontheorem} }
\def \bpr{ \begin  {proof} }
\def \epr{ \end  {proof} }
\def \ba{ \begin{algorithm} }
\def \ea{ \end{algorithm} }
\def \be{ \begin{example} }
\def \eex{ \end{example} }
\def \bes{ \begin{examples} }
\def \eexs{ \end{examples} }
\def \br{ \begin{remark} }
\def \er{ \end{remark} }
\def \brs{ \begin{remarks} }
\def \ers{ \end{remarks} }
\def \bpb{ \begin{problem} }
\def \epb{ \end{problem} }

\newcommand{\MP} {\mathrm{mp}}

\newcommand{ \GF} {\mathrm{GF}}

\newcommand{\lcm} {\mathrm{lcm}}

\newcommand{\ol} {\overline}

\newcommand{\F}{\mathcal{F}}

\begin{document}
\maketitle
\begin{abstract}
The Games-Chan algorithm finds the minimal period of a periodic binary sequence of period $2^n$, in $n$ iterations. We generalise this to periodic $q$-ary sequences  (where $q$ is a prime power) using generating functions and polynomials.
We apply this to find the multiplicity of $x-1$ in a $q$-ary polynomial $f$ in $\log_{\,q}\deg(f)$ iterations.
\end{abstract}


\section{Introduction}
In \cite{GC}, the authors gave a $\log_{\,2}N$ algorithm to find the minimal period of a periodic binary sequence with period $N=2^n$. Their proof used binomial coefficients $\bmod\ 2$ and combinatorial identities. We generalise \cite[Theorem]{GC} to periodic $q$-ary sequences of length $N=q^n$  (where $q$  is a prime power) using  generating functions  and $\GF(q)[x]$. We apply this to finding the multiplicity of $x-1$ in $f\in\GF(q)[x]$ in $\log_{\,q}\deg(f)$ iterations.

We conclude with the corresponding recursive, division-free algorithm.
\subsection{Notation}
Let $q=p^e$ for some prime $p$,  put $\F=\GF(q)$ and for $f\in\F$, write $|f|$ for $\deg(f)$.  We have $x^{p^r}-1=(x-1)^{p^r}$. Throughout the paper, $N=q^n$ for some $n>0$.
\subsection{Sequences} We write $s=(s_0,s_1,\ldots)$ for a  sequence with $s_i\in\F$. The {\em generating function} of $s$ is $\mathrm{G}(s)(x)=\sum_{i=0}^\infty\ s_i\ x^i\in\F[[x]]$. It is well-known that if $s$ has period $N$, we can write  
$$\mathrm{G}(s)(x)=\frac{\ol{s}(x)}{x^N-1}$$
 where $\ol{s}(x)=s_0+\cdots+s_{N-1}x^{N-1}\in\F[x]$ and $|\ol{s}|< N$.  The sequence $s$ has a unique monic {\em minimal polynomial}   $\mu(s)=(x^N-1)/\gcd(\ol{s},x^N-1)=(x-1)^{\MP(s)}=x^{\MP(s)}-1$; here $\MP(s)$ is called the {\em minimal period} of $s$. We have $\MP(s)=N-|\gcd(\ol{s},x^N-1)|$. The zero sequence has $\mathrm{G}(0)= 0/1$ and $\gcd(0,1)=1=(x-1)^0$ so $\MP(0)=0$.  If $t$ is another sequence, then
$\mu(s+t)=\lcm\{\mu(s),\mu(t)\})$, \cite[Theorem 8.59]{LN}.\\ 

The following notion of '$q$-folding' of a sequence plays a key role; it generalises the use of left and right halves of a sequence in \cite{GC}.
\bd \label{qfolding} Let  $s$ be a sequence with initial period $(s_0,\ldots,s_{N-1})\in\F^{N}$. Put  $s^{(i)}=(s_{iN'},\ldots,s_{(i+1)N'-1})\in\F^{N'}$ for $0\leq i\leq q-1$, so that $(s_0,\ldots,s_{N-1})=(s^{(0)},\ldots,s^{(q-1)})$. 
Also, define  $s^\bullet =s^{(0)}+\cdots+s^{(q-1)}\in\F^{N'}$ and $(s^{(0)})^\bullet=s^{(0)})$.
\ed

The main idea of the proof is to relate $\ol{s}$ and $\ol{s^\bullet}$.

\bt\label{mainthm} Let $s$ be a non-zero sequence over $\F$ of period $N$.  Then
$$(i)\ \ \mathrm{G}(s)=-\frac{ \ol{s^{(0)}} }{x^{N'}-1}-\cdots -\frac{x^{(q-2)N'}\ \ol{s^{(q-2)}} }{x^{(q-1)N'}-1}+\frac{x^{(q-1)N'}\ol{ s^\bullet}}{x^N-1}$$

 \begin{eqnarray*}
(ii)\ \MP(s)=\left\{\begin{array}{ll}
\max\{\MP(s^{(0)}),\ldots,\MP(s^{(q-2)})\} &\mbox {if }s^\bullet=0\\\\
\MP(s^\bullet)+(q-1)N' &\mbox{otherwise.}
\end{array}
\right.
\end{eqnarray*}
\et
\bpr We put   $\sigma=s^\bullet$  to simplify the notation.   Firstly, $\sigma=0$ if and only if $\ol{\sigma}=0$,  and if $\ol{\sigma}\neq0$ then $0\leq|\ol{\sigma}|< N'$. By definition 
$\ol{s^{(i)}}(x)=s_{iN'}+\cdots+s_{(i+1)N'-1}x^{(i+1)N'-1}$ and 
$$\ol{s}(x)=\ol{s^{(0)}}(x)+\cdots +x^{(q-1)N'}\ol{s^{(q-1)}}(x)=g_0(x)+\cdots +g_{q-1}(x)\ \mathrm{say,}$$ where $g_i(x)=x^{iN'}\ \ol{s^{(i)}}(x)$ for $0\leq i\leq q-1$. Thus
$x^{iN'}|g_i(x)$\,,
$$\ol{\sigma}(x)=g_0(x)+x^{-N'}g_1(x)+\cdots+x^{-(q-2)N'}g_{q-2}(x) +x^{-(q-1)N'}g_{q-1}(x)\,\ \mathrm{and}$$
\begin{eqnarray*}
\ol{s}&=& (g_0-x^{(q-1)N'}g_0)+\cdots +(g_{q-2} -x^{N'}g_{q-2})+(x^{(q-1)N'}g_0+\cdots+x^{N'}g_{q-2}+g_{q-1})\\
&=&(1-x^{(q-1)N'})g_0+\cdots +(1-x^{N'})g_{q-2}+x^{(q-1)N'}(g_0+\cdots+x^{-(q-1)N'}g_{q-1})\\
&=&(1-x^{(q-1)N'})g_0+\cdots +(1-x^{N'})g_{q-2}+x^{(q-1)N'}\ol{\sigma}
\end{eqnarray*}
so
$$\mathrm{G}(s)=-\frac{ \ol{s^{(0)}} }{x^{N'}-1}-\cdots -\frac{x^{(q-2)N'}\ \ol{s^{(q-2)}} }{x^{(q-1)N'}-1}+\frac{x^{(q-1)N'}\ \ol{\sigma}}{x^N-1}$$
and (i) is proved. Now  $\mu(\frac{x^{iN'}\ \ol{s^{(i)}} }{x^{(i+1)N'}-1})=\mu(s^{(i)})$ and  so 
$$\mu(s)=\lcm\{\mu(s^{(0)}),\ldots,\mu(s^{(q-2)}),\mu\left(\frac{x^{(q-1)N'}\ \ol{\sigma}}{x^N-1}\right)\}.$$
 So if   $\sigma=0$ then $\MP(s)=\max\{\MP(s^{(0)}),\ldots,\MP(s^{(q-2)})\}.$
Now let $\sigma\neq 0$.   For $0\leq i \leq q-2$, $\MP(s^{(i)})\leq|s^{(i)}|\leq {(i+1)N'-1}<(q-1)N'$, so $\max\{\MP(s^{(0)},\ldots,\MP(s^{(q-2)})\}$ $<(q-1)N'$. Also,   since $|\ol{\sigma}|<N'$,
$$N-|\gcd(x^{(q-1)N'}\ol{\sigma},x^N-1)|=qN'-|\gcd(\ol{\sigma},x^{N'}-1)|=(q-1)N'+\MP(\sigma)\geq(q-1)N'.$$ 
We conclude that  if $s^\bullet\neq0$ then $\MP(s)=\MP(s^\bullet)+(q-1)N'$.
\epr

\bc (\cite[Theorem]{GC}) Let $s$ be a non-zero sequence over $\GF(2)$ of period $N=2^n$.  Then
\begin{eqnarray*}
\MP(s)=\left\{\begin{array}{ll}
\MP(s^{(0)}) &\mbox {if }s^\bullet=0\\\\
\MP(s^\bullet)+N/2 &\mbox{otherwise.}
\end{array}
\right.
\end{eqnarray*}
\ec
\section{The multiplicity of $x-1$ in $f\in\F[x]$}
We next show how Theorem \ref{mainthm} yields a $\log_{\,q}$ division-free algorithm to find the multiplicty of $x-1$ in $f\in\F[x]$.

\bd For $f\in\F[x]^\times$, $\pi(f)$ denotes the multiplicity of $x-1$ in $f$. 
\ed

Suppose  $|f|< N$. Since $\gcd(f,x^N-1)=(x-1)^{\pi(f)}$, we can relate $\pi(f)$ and $\MP(f/(x^N-1)$: we have $\mu(f/(x^N-1))=(x^N-1)/\gcd(f,x^N-1)$, so $\MP(f/(x^N-1))=N-\pi(f)$ or $\pi(f)=N-\MP(f/(x^N-1))$.

To apply Theorem \ref{mainthm},  we first adapt the definition of $q$-folding to $\GF(q)[x]$.

\bd Let $f\in\F[x]^\times$, $N$ be the smallest power of $q$ greater than $|f|$ and put $f=f^{(0)}+\cdots+ f^{(q-1)}$ where $|f^{(i)}|< (i+1)N'$ and $x^{iN'}|f^{(i)}$ for $0\leq i\leq q-1$. We also put $f^{\bullet}=f^{(0)}+\cdots+ x^{-(q-1)N'}f^{(q-1)}$ (so that $|f^\bullet|<N'$) and $(f^{(0)})^\bullet=f^{(0)}$.
\ed

\bc\label{maincor} Let $f\in\F[x]^\times$ and $N$ be the smallest power of $q$ greater than $|f|$.  Then
 \begin{eqnarray*}
\pi(f)=\left\{\begin{array}{ll}
\min\{\pi(f^{(0)})+(q-1)N',\ldots,\pi(f^{(q-2)})+N'\} &\mbox {if }f^\bullet=0\\\\
\pi(f^\bullet) &\mbox{otherwise.}
\end{array}
\right.
\end{eqnarray*}
\ec
\bpr Put $s=f/(x^N-1)$. Then $\ol{s^{(i)}}=f^{(i)}$ for $0\leq i\leq q-1$ and $\ol{s^\bullet}=f^\bullet$, so from Theorem \ref{mainthm}, 
$$(i)\ \ \mathrm{G}(s)=-\frac{ f^{(0)} }{x^{N'}-1}-\cdots -\frac{x^{(q-2)N'}\ f^{(q-2)} }{x^{(q-1)N'}-1}+\frac{x^{(q-1)N'}\ f^\bullet}{x^N-1}\ \ \mathrm{and}$$

 \begin{eqnarray*}
(ii)\ \MP(s)=\left\{\begin{array}{ll}
\max\{\MP(s^{(0)}),\ldots,\MP(s^{(q-2)})\} &\mbox {if }s^\bullet=0\\\\
\MP(s^\bullet)+(q-1)N' &\mbox{otherwise.}
\end{array}
\right.
\end{eqnarray*}
If $f^\bullet=0$ then 
 \begin{eqnarray*}
\pi(f)&=& N-\max\{\MP(s^{(0)}),\ldots,\MP(s^{(q-2)})\}\\
&=&\min\{N-\MP(s^{(0)}),\ldots,N-\MP(s^{(q-2)})\}\\
&=&\min\{N+\pi(f^{(0)})-N',\ldots,N+\pi(f^{(q-2)})-(q-1)N'\}\\
&=&\min\{\pi(f^{(0)})+(q-1)N',\ldots,\pi(f^{(q-2)})+N'\}
\end{eqnarray*}
and 
$\pi(f)=N-\MP(f^\bullet/(x^N-1))=N+(\pi(f^\bullet)-N)$ otherwise.
\epr

\subsection{The Binary Case}
We give a direct proof of Corollary \ref{maincor} when $q=2$. This requires high school algebra and  $x^{2^r}+1=(x+1)^{2^r}$ in $\GF(2)[x]$ only.
\bp \label{binaryprop} Let $f\in\GF(2)[x]^\times$, let $N$ be the smallest power of 2 greater than  $|f|$  and $N'=N/2$. Then 
\begin{eqnarray*}
\pi(f)=\left\{\begin{array}{ll}
\pi(f^{(0)})+N' &\mbox {if }f^\bullet=0\\\\
\pi(f^\bullet) &\mbox{otherwise.}
\end{array}
\right.
\end{eqnarray*}
\ep
\bpr Put $g=f^{(0)}$ and $h=f^{(1)}$  to simplify the notation. We have 
$$f=(g+x^{N'}g)+(h+x^{N'}g)=(x+1)^{N'}g+x^{N'}(g+x^{-N'}h)=(x+1)^{N'}g+x^{N'}f^\bullet$$
so that if $f^\bullet =0$ then $\pi(f)=\pi(f^{(0)})+N'$. Now let $f^\bullet\neq0$. 
Firstly, $\pi(f^\bullet)\leq\pi(f)$ and $\pi(f^\bullet)\leq|f^\bullet|<N'$,  so that
\begin{eqnarray}\label{clincher}
\frac{f}{(x+1)^{\pi(f^\bullet)}}=(x+1)^{N'-\pi(f^\bullet)}g+\frac{x^{N'}\cdot f^\bullet}{(x+1)^{\pi(f^\bullet)}}.
\end{eqnarray}
If $\pi(f^\bullet)<\pi(f)$ then $x+1$ divides the left-hand side of Equation (\ref{clincher}). As $N'-\pi(f^\bullet)>0$, $(x+1)|(x+1)^{N'-\pi(f^\bullet)}g$. This implies that $(x+1)|x^{N'}$, which is a contradiction and therefore $\pi(f)=\pi(f^\bullet)$.
\epr
We can also deduce \cite[Theorem]{GC} from Proposition \ref{binaryprop} using the method of Corollary \ref{maincor}. The details are elementary and left to the reader.
\section{The $\MP$ Algorithm}
The corresponding  $\log_{\,q}N$ algorithm is division-free.\\

If $\ol{s}(1)=\sum_{i=0}^{|\ol{s}|}s_i\neq 0$ then $\MP(s)=N$, so we can include this in the corresponding algorithm, which requires $\log_2N$ iterations. 
\begin{tabbing}
{\tt procedure} $\MP(s; N$): {\tt return} $\MP(s)$;\\\\
\{\\
{\tt if} $(s=0)$  then $\MP:=0$; {\tt exit};\\
{\tt if}  $(\ol{s}(1)\neq0)$ then $\MP:=N$; {\tt exit};\\\\

{\tt if}\ ($N\geq 2$) {\tt then if} ($s^\bullet=0$) \= {\tt then} 
$\MP:= \max\{\MP(s^{(0)};N'), \ldots, \MP(s^{(q-2)};N');q-1\}$; \\

\>{\tt else} $\MP:=\MP(s^\bullet;N')+(q-1)N'$;\\
 \}
\end{tabbing}

\begin{tabbing}
{\tt procedure} $\max(n_1,\ldots,n_k;k):m$;\\
\{\\
if $k=1$ {\tt then} $m:=n_1$ {\tt else} $m:=\max\{n_1,\max\{n_2,\ldots,n_k;k-1\};2\}$;\\
\}              
\end{tabbing}

There is a corresponding multiplicity algorithm which is similar and left to the reader.

\bibliographystyle{plain}

\begin{thebibliography}{1}

\bibitem{GC}
\newblock{Games, R.A. and Chan, A.H.}
\newblock{A Fast Algorithm for Determining the Complexity of a Binary Sequence with period $2^n$}
\newblock{\em IEEE Trans. Inform. Theory}, 29,144--146, 1983.

\bibitem{LN}
\newblock{Lidl, R. and Niederreiter, H.}
\newblock{\em Finite Fields.}
\newblock{Encyclopedia of Mathematics and its Applications, Vol. 20. Addison-Wesley Co., Reading MA,1983.}

\end{thebibliography}

\end{document}